\documentstyle[sprocl,psfig]{article}
\def\be{\begin{equation}}
\def\ee{\end{equation}}
\def\bea{\begin{eqnarray}}
\def\eea{\end{eqnarray}}

\begin{document}

\title{SPECTROSCOPIC GRADIENTS IN EARLY -- TYPE GALAXIES AND IMPLICATIONS
ON GALAXY FORMATION}

\author{D. Mehlert, R. Bender, R. P. Saglia}
\address{Universit{\"a}tssternwarte M{\"u}nchen, Scheinerstra{\ss}e 1, 
D 81679, M{\"u}nchen\\E-mail: mehlert/bender/saglia@usm.uni-muenchen.de} 
\author{G. Wegner}
\address{Department of Physics and Astronomy,
Dartmouth College,\\
NH 03755-3528, USA}

\maketitle\abstracts{The Coma cluster is the ideal place to study 
galaxy structure
as a function of environmental density in order to constrain theories 
of galaxy formation and evolution. Here we present the spectroscopy 
of 35 early type Coma galaxies, which shows that 
the age spread of early type galaxies in the Coma cluster is large (15 Gyrs).
In contrast to the field, 
the dominant stellar population in all (massive)
Coma Es is older than 8 Gyr, while
only S0s, which possess extended disks, can be as young as 2 Gyr. 
The old, most massive Es show a strong light element 
enhancement, probably due to a rather short star formation time scale 
and hence to a SNII -- dominated element enrichment. 
The lower mass S0s are much less enhanced in light elements,
indicating a longer star formation time scale.
The measured absorption line index gradients support the idea that
early type galaxies formed in processes that include
both stellar merging and gaseous dissipation.
}

\vspace*{-5mm}

\section{The Project}

We obtained long slit spectra for 
35 E and S0 galaxies at the Calar Alto 3.5 m, 
the Michigan -- Dartmouth -- M.I.T. 2.4 m and the McDonald 2.7 m telescopes.
Our sample is complete for galaxies brighter than
M$_B$ = -21.6$^{mag}$ (14 galaxies) and is 1/3 complete
in the magnitude range -21.6$^{mag}$ $< M_B <$ -20.2$^{mag}$ (21 galaxies). 
The integrated exposure time per galaxy ranges from 2 -- 5 h.
Our sample spans 4 dex of the cluster density reaching 4 Mpc out
of the center.
We derived the {\it radial profiles}
of the line indices Mg, Fe and H$_{\beta}$ as well as of the stellar rotation 
and the velocity dispersion reaching out to 2 r$_e$ for most of the galaxies. 

\vspace{-2mm}

\section{Age and Metallicity distribution}

\begin{figure}[t]
\psfig{figure=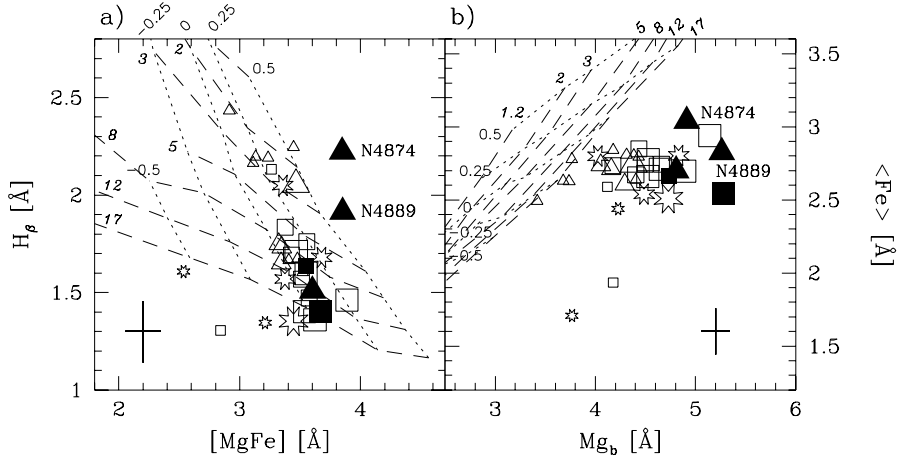,height=6.0cm}
\caption{{\footnotesize Mean absorption line indices inside 0.5 r$_e$
for 35 Coma galaxies compared to the model grids of Worthey
(1994); dashed lines - constant age t in Gyr; dotted lines  -
constant metallicity log(Z/Z$_{\odot}$).
Open squares represent Es, open triangles S0 and open stars E/S0s, while black
triangles represent the three CD galaxies (NGC 4874, NGC 4889, NGC 4839)
and black squares two galaxies (NGC 4816 \& IC 4051),
which contain kinematically decoupled/peculiar cores (Mehlert et al. 1997).
Different sizes refer to small, medium and large mean velocity dispersion
$\sigma$ - or mass - inside 0.5 r$_e$.
{\bf (a)} The age indicating index H$_{\beta}$ versus
the metallicity indicating $[$MgFe$]$ combined index. The 
cross indicates the mean index errors for
all galaxies. These translate into errors in 
metallicity 
$\Delta$log(Z/Z$_{\odot}$) $\approx$ 0.25 dex and errors in age 
$\Delta$t $\approx$ 2 -- 9 Gyr for young and old
galaxies, respectively.
{\bf (b)} $<$Fe$>$ versus the Mg$_b$ index. The mean index errors for
all galaxies are indicated by the cross.}}
\label{fig:agemet}
\vspace*{-4.1mm}
\end{figure}

We used Worthey's (1994) stellar population synthesis models  
to investigate the age and metallicity distribution of our sample. 
In Fig. \ref{fig:agemet}a we 
plotted the grid of models for the age sensitive H$_{\beta}$ index versus 
the metallicity sensitive 
combined index $[$MgFe$]$ = (Mg$_b$ $<$Fe$>$)$^{1/2}$, where 
$<$Fe$>$ = (Fe$_{5270}$ + Fe$_{5335}$)/2. Overplotted are
the mean (inside 0.5 r$_e$) indices we measured for our galaxies.
According to Worthey's models the ages of our Coma galaxies range
from 2 -- 17 Gyr. However, all massive Es are
older than 8 Gyr having a mean age of $\approx$ 12 Gyr. 
Interestingly the two cD galaxies of the main cluster
(NGC 4874 \& NGC 4889) are younger than the Es, but the most metal rich
galaxies in Coma. Only the 
less massive S0s, which possess an extended disk component, can be as 
young as 2 Gyr, but also as old as 12 Gyrs. Similar results have been found
by Kuntschner \& Davies (1997, KD97) in the Fornax
cluster, where the central number density of early -- type galaxies is 6
times lower than in the Coma cluster.
In contrast field 
Es are all younger than 10 Gyr and can also be as young as 2 Gyr 
(Faber et al. 1995). 
Similar differences between the effective age of the stellar 
populations in field and cluster Es have been found by Rose et al. (1994)
as well as Guzman et al. (1992).
We detect no dependence of a galaxy's age or 
metallicity on the density profile of the Coma cluster. 
The small metallicity spread (0.25 dex) of our Coma sample is probably only
due to the lack of faint galaxies in our sample (M$_B >$ -20.2$^{mag}$). 
For example, 
KD97 found a metallicity spread of $\approx$ 0.75 dex
for their complete sample of 22
Es and S0s in the Fornax cluster reaching M$_B$ = -17$^{mag}$.\\
It is important to emphasize that the ages and metallicities we derived
depend strongly on the stellar population model 
we used. 
An obvious problem is the Mg$_b$ to $<$Fe$>$ overabundance.
Fig. \ref{fig:agemet}b shows the 
mean $<$Fe$>$ versus the Mg$_b$ index measured for our 
galaxies and the grid of Worthey's models. 
Obviously, the models {\it do not} reproduce
the data in the sense that the galaxies are highly 
overabundant in Mg$_b$ relative to $<$Fe$>$. The most massive Es show 
abundances up to 0.5 dex super solar, while less 
massive S0s show solar element ratios or an enhancement up to 0.25 dex 
solar. These properties are also known for field Es (FFI95;
G93). 
The supersolar light element enhancement of massive Es is probably due to
a SN II dominated element enrichment. 
In a recent study Thomas et al. (1997) show that in addition to a top heavy 
initial mass function (IMF) a short star formation time scale is needed to 
produce a high SN II/SNIa ratio and hence the enhancement, which is 
detected in the most 
massive early type galaxies. As long as the available stellar population 
models cannot reproduce the observed abundances 
the determination of the ages and 
metallicities of stellar populations still remain problematic.

\section{Line strength gradients}

\begin{figure}
\psfig{figure=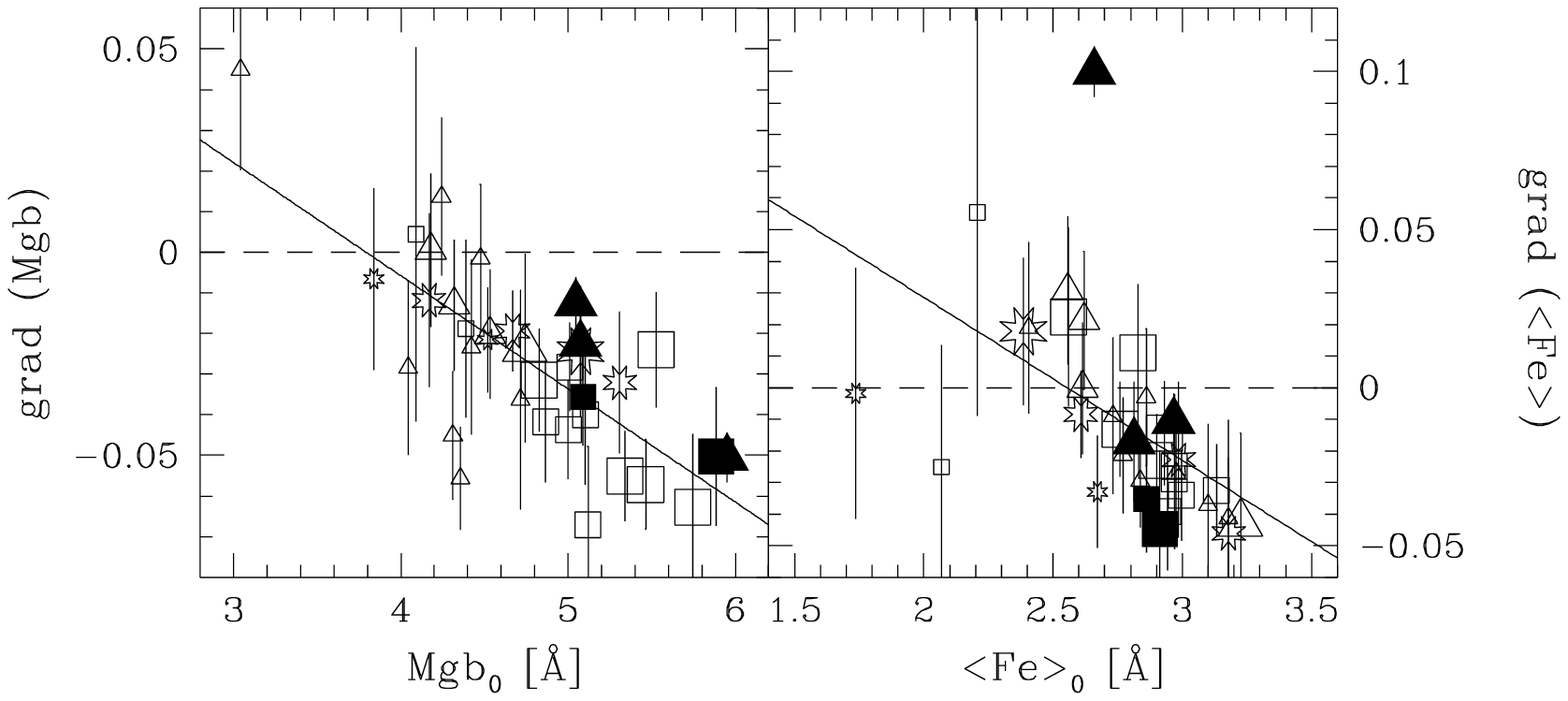,height=5.3cm}
\caption{{\footnotesize Fitted Mg$_b$ {\bf (a)} and $<$Fe$>$ {\bf (b)}
gradients versus the
central fit value at 0.01 r$_e$ for 35 Coma early type galaxies (symbols and
sizes as in Fig. \ref{fig:agemet}). A linear ${\chi}^2$ fit 
yields following
correlations: grad(Mg$_b$) = 0.106 - 0.028 Mgb$_0$
with an RMS scatter of $\sigma$ = 0.0044 in
grad(Mg$_b$) and
grad($<$Fe$>$) = 0.132 - 0.052 $<$Fe$>_0$
with an RMS scatter of $\sigma$ = 0.0142 in
grad($<$Fe$>$). The probability of having no correlation can be excluded
at the 
3~$\sigma$ and 2.7~$\sigma$ levels for the Mg$_b$ and $<$Fe$>$ gradients,
respectively.
}}
\label{fig:grad_cent}
\end{figure}

We computed index gradients 
grad(index) = $\Delta$log(index)/$\Delta$log(r/r$_e$)
by applying a linear 
${\chi}^2$ fit to the index profiles out to r$_e$.
In Fig. \ref{fig:grad_cent}a \& b we 
plotted the Mg$_b$ and $<$Fe$>$ gradients versus 
the central fit value at 0.01\,r$_e$. 
The most massive Es
show the highest central indices and the strongest metallicity
gradients, while
the less massive S0s have lower central values and flatter gradients.
The gradients
of the age sensitive H$_{\beta}$ index are in the range of
grad(H$\beta ) = - 0.18 \mbox{ to } 0.07$.
Together, these considerations support the idea that early -- type galaxies 
formed by a combination of merging and dissipative (gas involving) processes.
Similar results have been found for field Es (Gonzales \& Gorgas 1995; 
Carollo et al. 1993).\\
Once Fig. \ref{fig:grad_cent} is taken into account, 
the Mg$_b$
gradient tends to be larger (on the 1~$\sigma$ level) for galaxies at larger 
distances from the cluster's center. 
A larger sample is needed to investigate the reality of this possible
environmental effect.
\vspace{0.5cm}

\noindent {\bf Acknowledgments} {\footnotesize
The authors thank the staff of the Calar Alto, MDM and McDonald observatory 
for their effective support. We thank
Laura Greggio, Daniel Thomas and Bodo Ziegler for helpful discussions.
This work was supported by the Deutsche Forschungsgemeinschaft via 
project Be 1091/6.}\vspace{1cm}\\
\noindent {\bf References}\\
\noindent \footnotesize{
Carlberg, R., 1984, ApJ, 286, 404\\
Carollo, C.M., Danziger, I.J., Buson, L., 1993, MNRAS, 265, 533\\
Davies, R.L., Sadler, E.M., Peletier, R.F., 1993, MNRAS, 262, 650\\
Faber, S.M., Trager, S.C., Gonzales, J.J., Worthey, G., 1995, IAU Symp. 164\\
\hspace*{0.4cm}({\it Stellar Populations}), eds. P.C. van der Kruit \& G.
Gilmore, p. 249\\
Fisher, D., Franx, M., Illingworth, G., 1995, ApJ 448, 119\\
Gonzalez, J.J., 1993, PhD Thesis, University of California, Lick Observatory\\
Gonzalez, J.J., Gorgas,J., 1995, in ASP Conference Series 86 
({\it Fresh Views on Elliptical}\\
\hspace*{0.4cm}{\it Galaxies}), eds. Buzzoni, A., Renzini, A., Serrano, A.,
p. 225\\
Guzman, R., Lucey, J. R., Carter, D., \& Terlevich, R. J., 1992, MNRAS, 257, 187\\
Kuntschner, H., Davies, R.L., 1997, MNRAS, submitted\\
Larson, R.B., 1976, MNRAS, 176, 31\\
Mehlert, D., Saglia, R.P., Bender, R., Wegner, G., 1997, A\&A, in press\\
Rose, J.A. Bower, R.G., Caldwell, N., Ellis, R.S., Sharples, R.M., Teague,
P., 1994, AJ, 108,\\
\hspace*{0.4cm}2054\\
Thomas, D., Greggio, L., Bender, R., 1997, MNRAS, in press\\
Worthey, G., 1994, ApJS, 95, 105\\
}
\end{document}